\documentclass[runningheads]{llncs}
\usepackage{graphicx}
\usepackage[table]{xcolor}
\usepackage{multirow}
\usepackage{booktabs}
\usepackage{adjustbox}
\usepackage{upgreek}
\usepackage{caption}
\usepackage{subcaption}
\usepackage{amsmath}

\usepackage{xcolor}

\newcommand{\myfigref}[1]{Fig. \ref{#1}}
\usepackage{comment}

\begin{document}
\title{Dictionary-based Method for Vascular Segmentation for OCTA Images}
\titlerunning{Dictionary-based Segmentation}
%

\author{Astrid M. E. Engberg \and
Vedrana A. Dahl \and
Anders B. Dahl}

\authorrunning{A. Engberg et al.}
%
\institute{Technical University of Denmark, Kgs.~Lyngby, Denmark
\email{\{asteng,vand,abda\}@dtu.dk}}

\maketitle  
%

%
%
\begin{abstract}
    Optical coherence tomography angiography (OCTA) is an imaging technique that allows for non-invasive investigation of the microvasculature in the retina. OCTA uses laser light reflectance to measure moving blood cells. Hereby, it visualizes the blood flow in the retina and can be used for determining regions with more or less blood flow. OCTA images contain the capillary network together with larger blood vessels, and in this paper we propose a method that segments larger vessels, capillaries and background. The segmentation is obtained using a dictionary-based machine learning method that requires training data to learn the parameters of the segmentation model. Here, we give a detailed description of how the method is applied to OCTA images, and we demonstrate how it robustly labels capillaries and blood vessels and hereby provides the basis for quantifying retinal blood flow.
\end{abstract}

\section{Introduction}
Optical coherence tomography angiography (OCTA) is a relatively novel imaging method commercially available in 2014 \cite{spaide2018optical}. Compared to alternative retinal imaging modalities, it is fast and provides high-resolution, depth-resolved images of the retinal microvasculature without any invasive procedures such as contrast agents \cite{kashani2017optical}. Despite the rapid acquisition, images can easily be corrupted by motion artefacts and noise. Noise induced by a low signal-to-noise ratio can occur due to eye conditions such as cataract \cite{spaide2018optical}, where the laser light, that illuminates the retina as part of the OCTA imaging system, is scattered by the cataract-affected lens.
Furthermore, the image intensity in OCTA images can vary over the image plane and give rise to bias in the image. These effects must be accounted for when choosing a method for obtaining an automatic segmentation of the retinal microvasculature.

\begin{figure}[tb]
    \centering
    \begin{subfigure}{0.45\textwidth}
        \includegraphics[width=\textwidth]{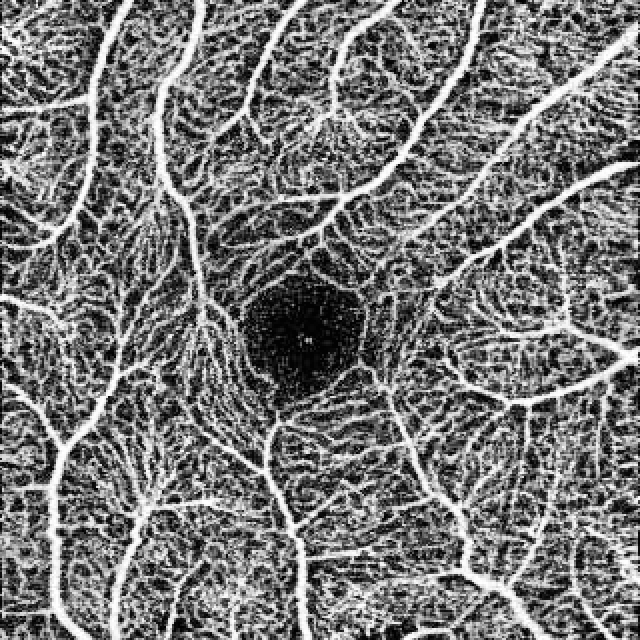}
        \caption{}
        \label{fig_sub:srl}
    \end{subfigure}
    ~\hspace{0.3cm}
    \begin{subfigure}{0.45\textwidth}
        \includegraphics[width=\textwidth]{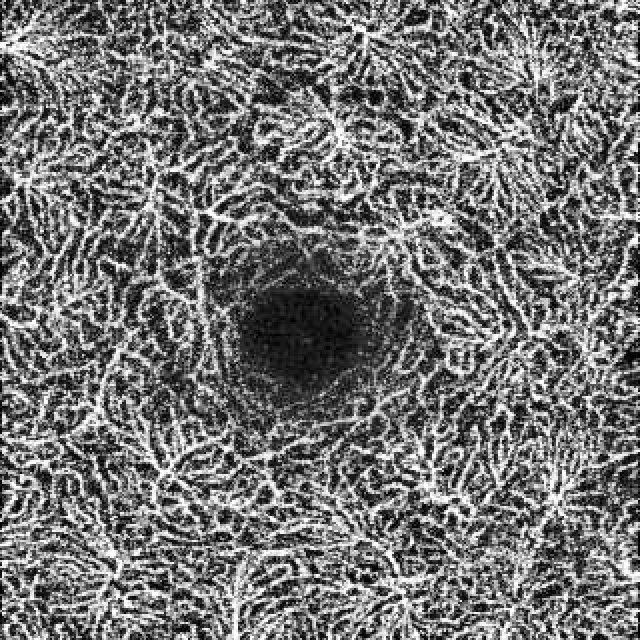}
        \caption{}
        \label{fig_sub:drl}
    \end{subfigure}
    \caption{Example of an OCTA image including (a) the superficial retinal layer (SRL) and (b) the deep retinal layer (DRL).} \label{fig:OCTA_example}
\end{figure}

OCTA imaging is a 3D acquisition method \cite{kashani2017optical}, but the clinical scanners employed in our studies perform a preprocessing of the data that segments the volume into so-called en face angiograms of different retinal layers. We focus on two layers: the superficial retinal layer (SRL) and the deep retinal layer (DRL). An example is shown in \myfigref{fig:OCTA_example}. Here, the larger blood vessels (arterioles and venules) are seen as thicker bright structures and the capillary network is a finer network of bright structures between the larger vessels. The dark area in the middle is the foveal avascular zone (FAZ). 

The problem we address is to segment OCTA images into three classes including larger vessels, capillaries, and background. By distinguishing between larger vessels and capillaries, the two structures can be analyzed separately. This allows for removing the influence of the size of the larger vessels, when quantifying the retinal capillaries, and hence not overestimating their density. We will solve this as a pixel labeling problem, such that we assign each pixel to one of three labels using a dictionary-based segmentation method.

The majority of clinical studies focus on solely obtaining quantitative metrics of the microvasculature, and segmentation of the microvasculature from OCTA images is a problem that has been addressed in only a few studies. Most studies obtain a segmentation through thresholding and filtering schemes \cite{chu2016quantitative,kim2016quantifying}. A few studies utilize manually annotated data to create segmentation models, such as probabilistic models \cite{eladawi_automatic_2017}, convolutional neural networks \cite{prentasic_segmentation_2016}, and Hessian- and deep learning-based methods \cite{deng_measurements_2018} to segment all vessels. A single study \cite{deng_measurements_2018} automatically segments main vessels and capillaries separately in retinal images using deep learning. 

One anatomical difference between larger vessels (arterioles and venules) and capillaries is their thickness. While the diameter of the capillaries is around 4-9 $\upmu$m \cite{kashani2017optical} and is determined by the size of the red blood cells, the larger vessels are thicker than capillaries and they vary more in size. Since the larger vessels and capillaries are connected, it is not trivial to design a model that separates the two anatomical structures. It cannot be accomplished by a simple thresholdning method, which are commonly applied to OCTA data.  
Instead, we propose to use the dictionary-based segmentation method from \cite{dahl2011learning,dahl2014dictionary,dahl2015dictionary}, where the segmentation model is learned from annotated training data. We have used this method for segmenting retinal microvasculature from OCTA images in \cite{engberg_automated_2020,engberg2020mia,engberg2019automated}. 

We introduced the fundamentals of the dictionary-based segmentation method in \cite{dahl2011learning}. It has later been extended for efficient computation of label probabilities which allowed iteratively updating label probabilities, which we used for computing deformable boundary models in \cite{dahl2014dictionary,dahl2015dictionary}, where probabilities are computed in each iteration. Further, we extended the model to allow for interactive segmentation in \cite{dahl2018content}, which allows for computing label probabilities from partially annotated data. 

In this paper we will focus on details related to segmenting microvasculature from OCTA images. Core elements of the method are described in e.g.~\cite{dahl2018content}, but to give a complete description of the method we will also describe them here. Furthermore, we use a feature-based representation to characterize local texture instead of using intensity patches. This has not previously been described, so we will provide the details here.

\begin{figure}[!tb]
    \centering
        \includegraphics[width=\textwidth]{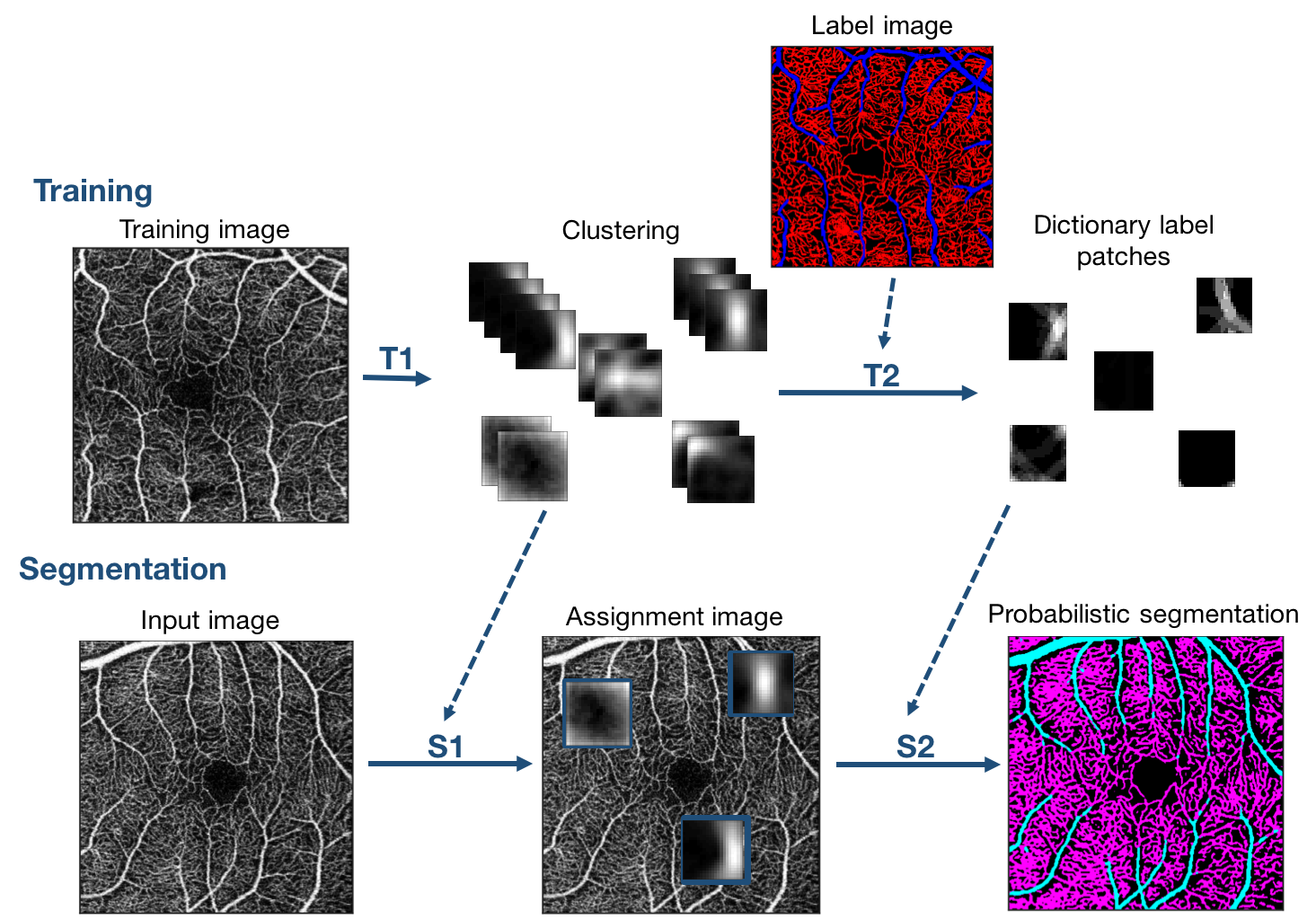}
    \caption{Illustration of the pipeline for the dictionary segmentation method. It consists of a training and a segmentation part. The training is based on a training image with a corresponding label image. Here, the training labels are blue for larger vessels, red for capillaries, and black for background. Based on patches sampled in the training image we perform a clustering (T1), which makes up the dictionary. Dictionary label patches are computed from the clustering of the training image and the label image (T2). In the segmentation part, we assign the dictionary to an input image (S1) and then compute pixel-wise probabilities of the labels using the dictionary labels (S2). Finally, we obtain the segmentation illustrated here. Larger vessels are marked with cyan, capillaries are magenta, and background is black. 
    }
    \label{fig:pipeline}
\end{figure}

\section{Method}
The basic principle of the dictionary-based segmentation model was introduced in \cite{dahl2011learning}, which is inspired by sparse coding methods \cite{elad2010sparse}. Sparse coding methods operate on image patches, and were originally applied to problems like denoising and texture modeling. Our dictionary-based segmentation method is similar to sparse coding methods because it employs a dictionary of image patches. However, here we assign each patch to only one dictionary element, which is different from sparse coding, where an image patch is typically represented by a small number of dictionary patches.

The idea of our method is that image patches with similar appearance should have the same label. We exploit this idea by clustering image patches (unsupervised part of training), computing patch label information from user input on the training image (supervised part of training), and then pasting this information in a testing image (using the model).
An overview of our segmentation pipeline is shown in \myfigref{fig:pipeline}.



\subsection{Training the model}
For training the segmentation model we need training data, consisting of image data and user-provided labeling. The image data used for training needs to be representative of the segmentation problem. This is usually one image or a small set of images. User-provided labeling should provide the desired segmentation for the images. It is not a requirement that all image data used for training is labeled by the user, but labeling should cover the variability of the structures to be segmented. 

\subsubsection{Extracting patch vectors. } 
We aim to construct a feature descriptor characterizing the local appearance around every image pixel. We start by extracting an $N \times N$ patch around the pixel. We choose $N$ to be odd such that the patch can be centered on a pixel, and we rearrange the patch into a vector of length $N^2$.

\subsubsection{Reducing the dimensionality of the patch vectors. }
We preform principal component analysis (PCA) to reduce the dimensionality of the patch vectors. Here, we randomly select $K$ patch vectors of length $N^2$, denoted $v_k$. For this set of vectors we compute a mean $\bar{v}$, such that we can compute patch vectors centered in origo $v_k-\bar{v}$, and arrange those vectors in the rows of a matrix $\mathbf{V}$, which will have size $K \times N^2$.

Now, we perform eigendecomposition of a matrix
\begin{equation}
    \mathbf{U} = \mathbf{V}^\mathrm{T} \mathbf{V} \ ,
\end{equation}
and keep eigenvectors corresponding to $q$ largest eigenvalues in a $N^2 \times q$ matrix $\mathbf{S}$. 

For each image pixel $i$ and its patch vector, we can now compute the projection
$f_i = (v_i-\bar{v})^\mathrm{T} \mathbf{S}$, and use it as a feature vector of length $q$.

\subsubsection{Incorporating image derivatives. } 
\begin{figure}[!tb]
    \centering
    \begin{subfigure}{0.28\textwidth}
        \includegraphics[width=\textwidth]{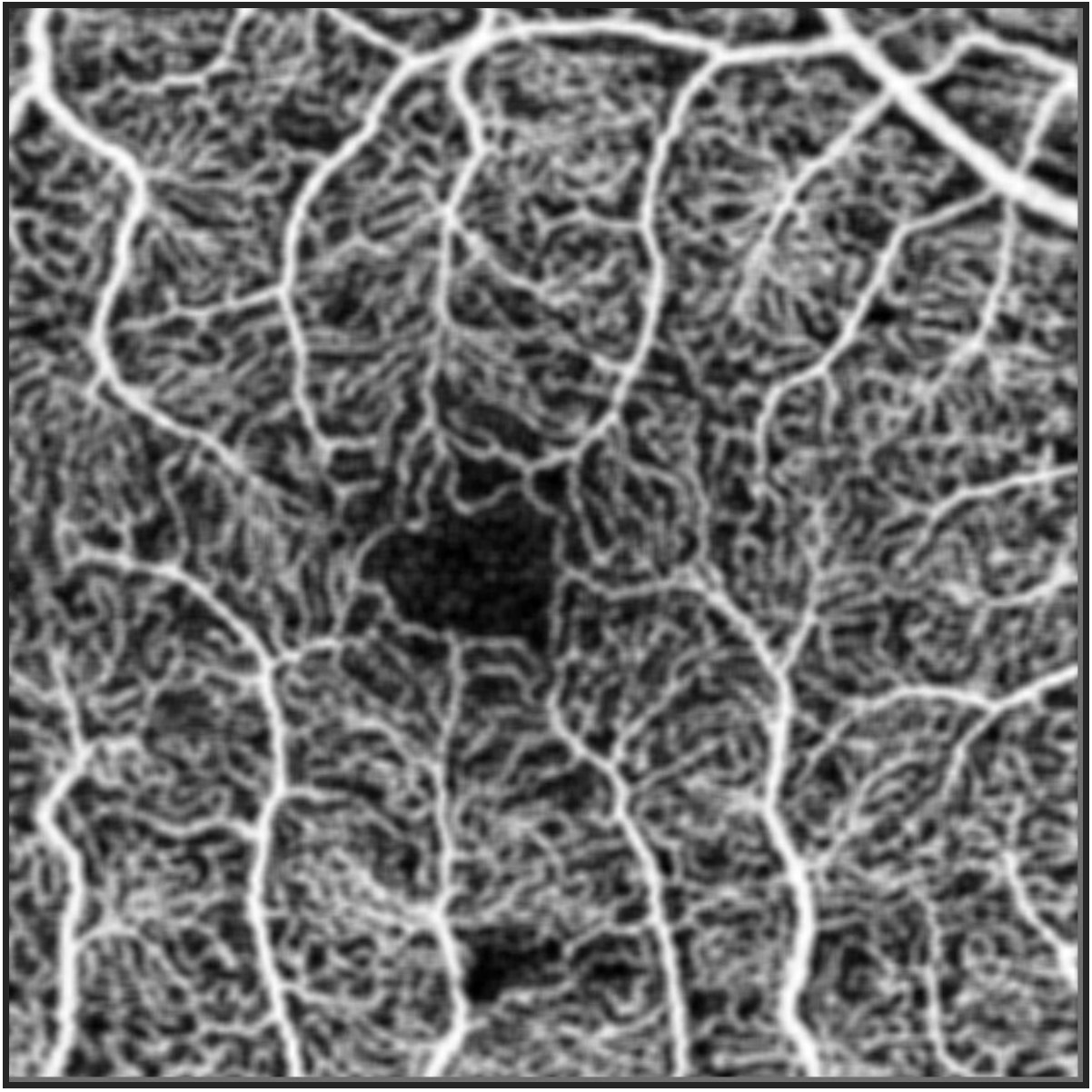}
        \caption{$I$}
    \end{subfigure}
    ~\hspace{0.3cm}
    \begin{subfigure}{0.28\textwidth}
        \includegraphics[width=\textwidth]{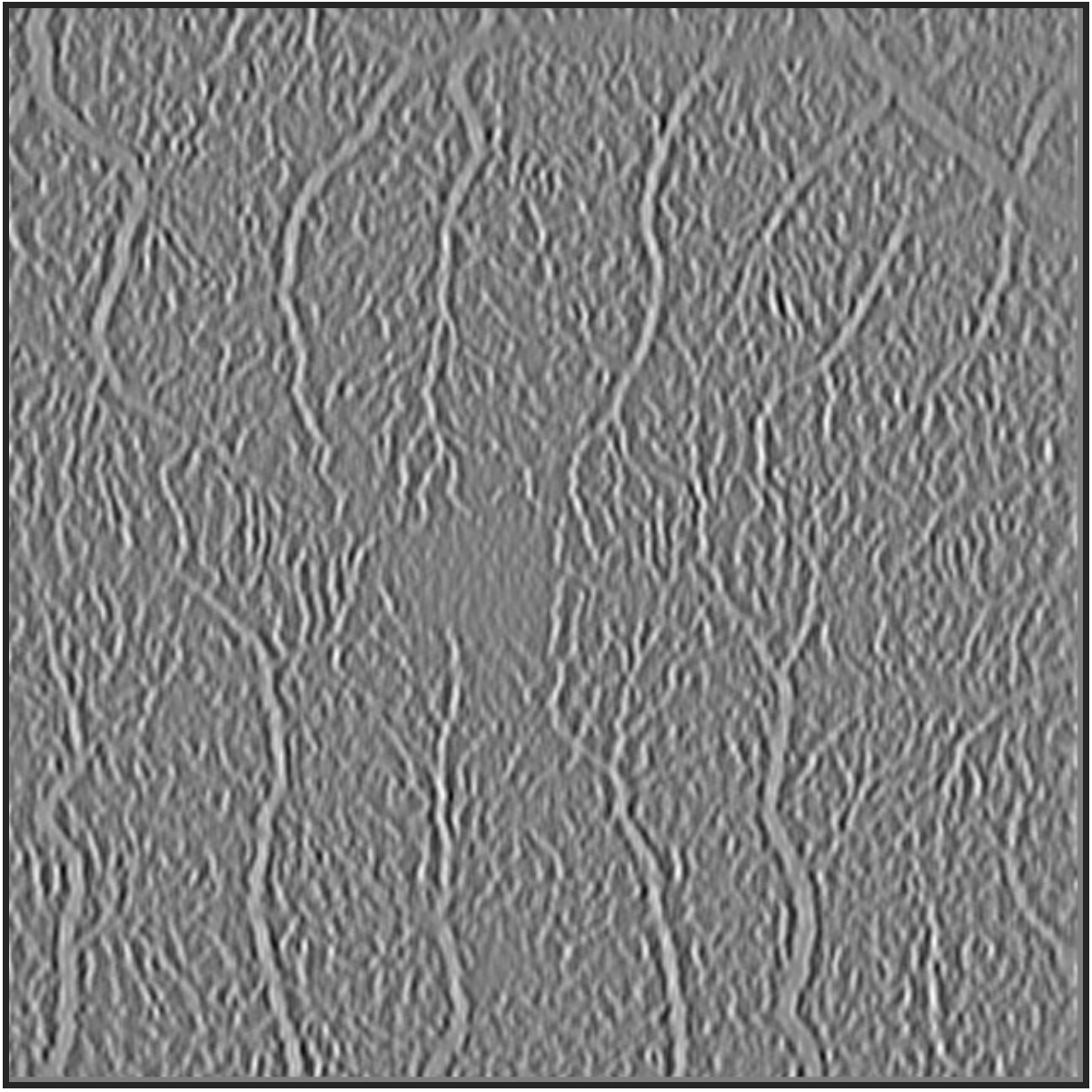}
        \caption{$I_{x}$}
    \end{subfigure}
        ~\hspace{0.3cm}
    \begin{subfigure}{0.28\textwidth}
        \includegraphics[width=\textwidth]{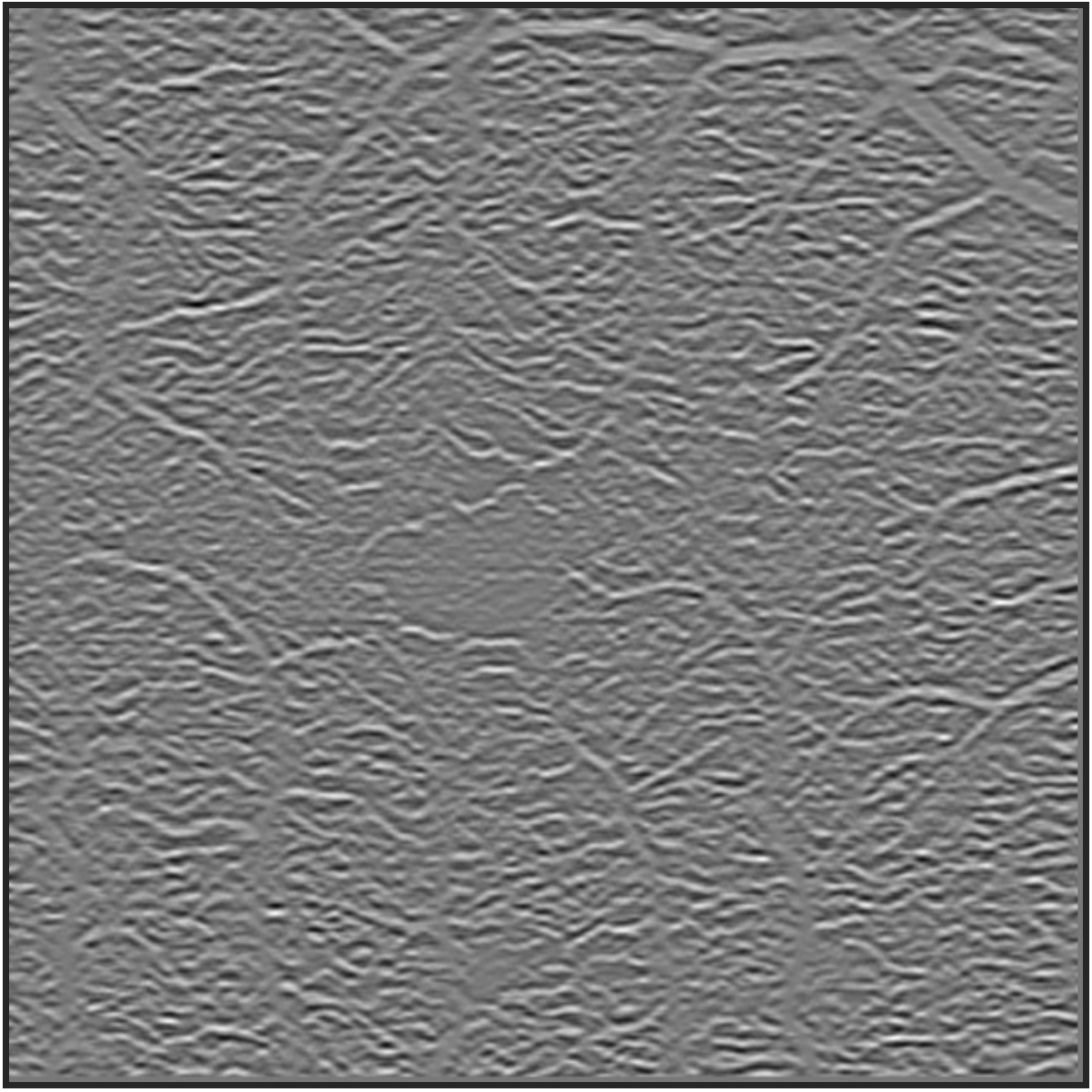}
        \caption{$I_{y}$}
    \end{subfigure}
    ~\hspace{0.3cm}
    \begin{subfigure}{0.28\textwidth}
        \includegraphics[width=\textwidth]{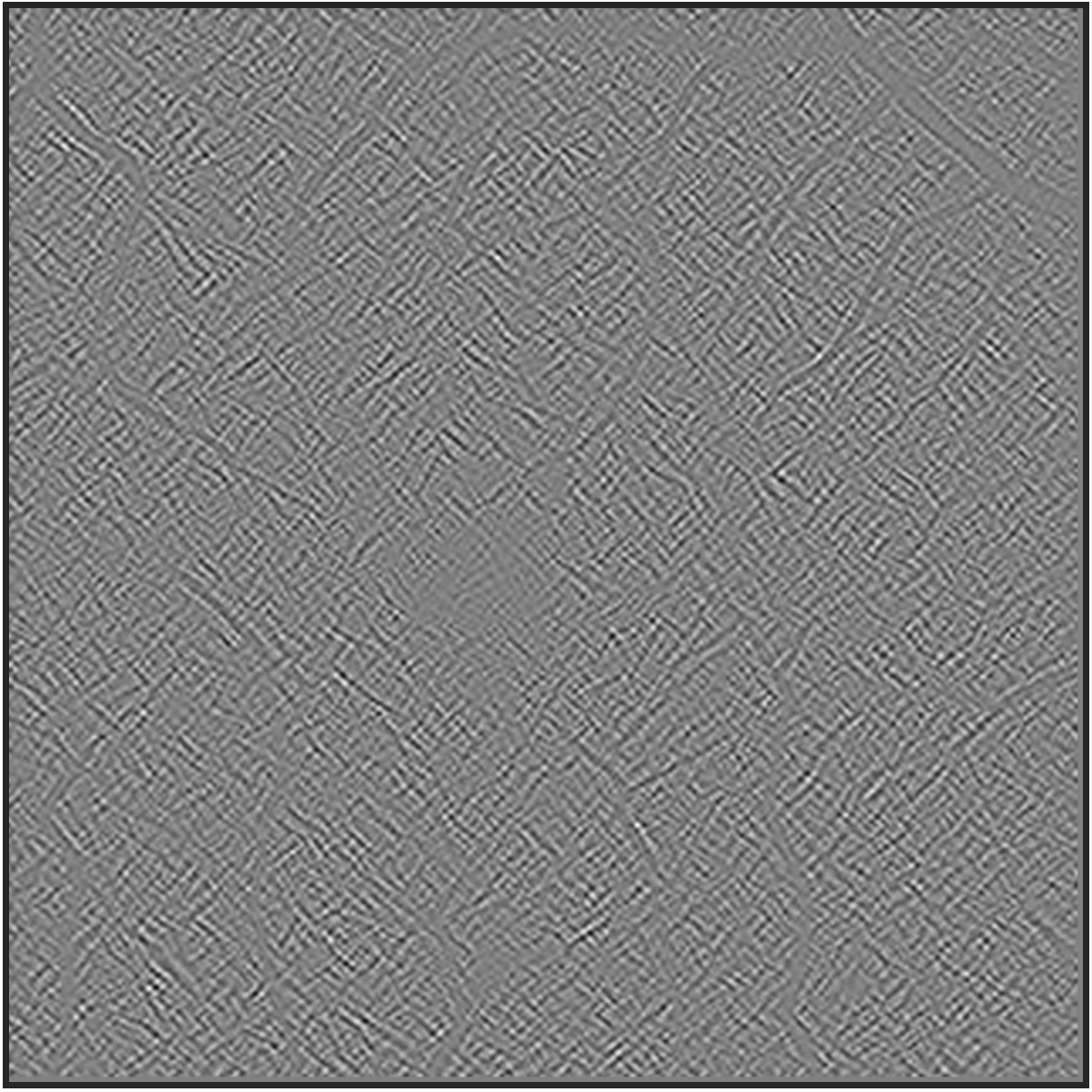}
        \caption{$I_{xy}$}
    \end{subfigure}
    ~\hspace{0.3cm}
    \begin{subfigure}{0.28\textwidth}
        \includegraphics[width=\textwidth]{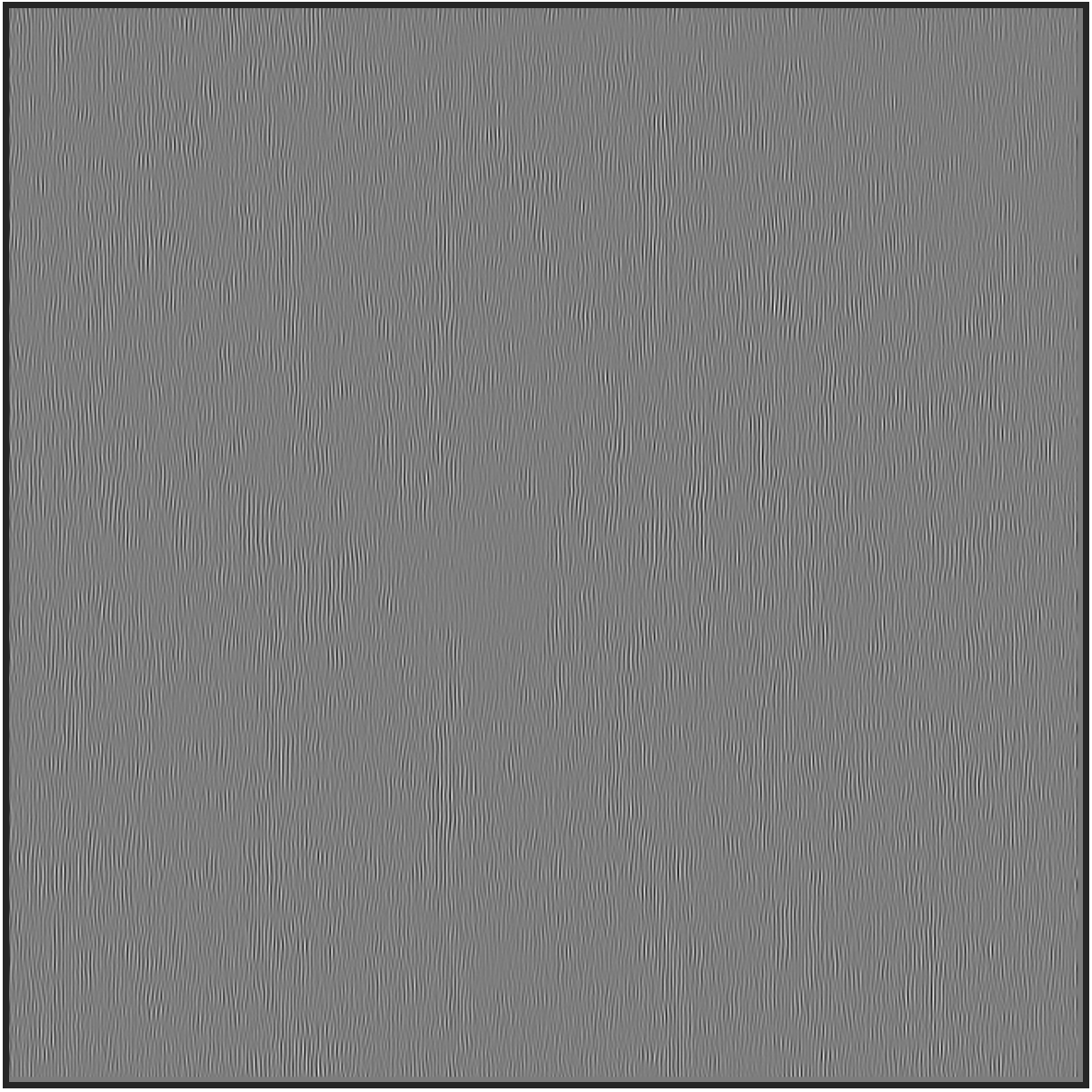}
        \caption{$I_{xx}$}
    \end{subfigure}
        ~\hspace{0.3cm}
    \begin{subfigure}{0.28\textwidth}
        \includegraphics[width=\textwidth]{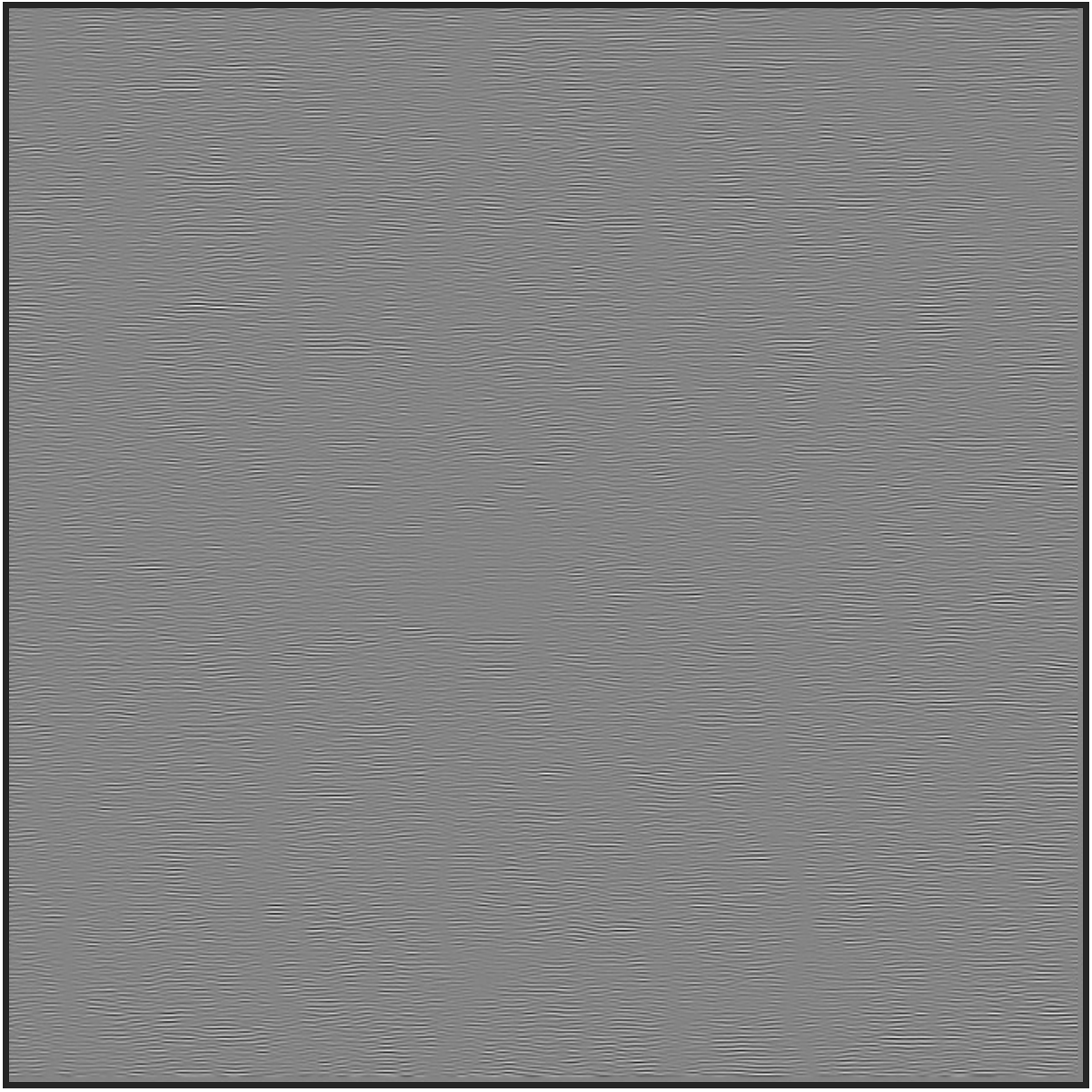}
        \caption{$I_{yy}$}
    \end{subfigure}
    \caption{Subfigures showing the component with the largest variability of the local features of the raw intensities, $I$, and the first ($I_{x}$, $I_{y}$) and second order derivatives ($I_{xy}$, $I_{xx}$, $I_{yy}$) with a feature patch size of $7 \times 7$ pixels.}
    \label{fig:comp}
\end{figure}

We strengthen the features employed in our method by incorporating the values of the first and second derivatives of the image
\begin{equation}
    I_x = \frac{\partial I}{\partial x} \ , \
    I_y = \frac{\partial I}{\partial y} \ , \
    I_{xx} = \frac{\partial^2 I}{\partial x^2}\ , \
    I_{xy} = \frac{\partial^2 I}{\partial x \partial y}\ , \
    I_{yy} = \frac{\partial^2 I}{\partial y^2} \ .
\end{equation}
For each of these five images we follow the procedure as described for the intensity image $I$, i.e. patch vector extraction and dimensionality reduction using PCA. This results in five additional feature vectors of length $q$. When all these are concatenated, we are left with a $6q$ feature vector per every image pixel. The PCA feature vector describes the local appearance of the image around the pixel. \myfigref{fig:comp} illustrates the PCA features computed in an OCTA image.

\subsubsection{Clustering. } 
The PCA feature vectors are clustered using a $k$-means hierarchical clustering to obtain the dictionary. As a distance measure, we use the Euclidean distance between PCA feature patches.

Hierarchical clustering is chosen instead of a conventional $k$-means because it speeds up the clustering dramatically. This allows for large dictionaries, and gives a very efficient search structure for assigning new feature vectors to the dictionary. 

The $k$-means hierarchical clustering is governed by two parameters, a branching factor $b$ and a depth $t$. A set of feature vectors is clustered by first clustering all vectors to $b$ clusters using conventional $k$-means, and then clustering each of these groups into $b$ sub-clusters. This is repeated $t$ times or until there are less than $b$ feature vectors in a cluster. The result is a tree graph, the $k$-means tree, where the nodes represent cluster centers. 

For our purposes, clustering is performed on PCA feature vectors corresponding to patches extracted from the training image. We therefore expect that features belonging to the same cluster correspond to image patches which have a similar appearance. 

\subsubsection{Incorporating user labelings. }

As mentioned previously, we expect similar image features to have similar labels. Having computed clusters of the image features, we need to define a labeling for each cluster. For this we need user labelings. This step constitutes the supervised part of building the dictionary.

For an image of the size $n \times m$, user labelings are stored as an $n \times m \times C$ array $\mathbf{L}$, where $C$ is the number of labels. In our case $C$ is three since we are interested in larger vessels, capillaries and background. Elements of $\mathbf{L}$ are binary, such that each pixel position belongs to only one label, indicated by a value $1$ in one layer of $\mathbf{L}$.

We now compute labeling information for each cluster in the dictionary by combining labeling information of all the members of the cluster. For this we extract patches of size $M \times M \times C$ from $\mathbf{L}$, where $M$ is chosen to be odd such that patches are centered on a pixel.

For each dictionary cluster, we extract such patches at the same spatial locations as the image patches belonging to the cluster. This set of patches is then averaged, and we obtain labeling information for each dictionary cluster. Due to the averaging of binary values, labeling information associated with dictionary patches is not binary, but can rather be interpreted as probabilities.

\subsection{Using the model}
Once we have a dictionary and the dictionary probabilities, we can use our model to process a new image (a testing image) and obtain pixel-wise probabilities of belonging to each of $C$ labels. 

\subsubsection{Dictionary assignment.} 
\begin{figure}[!tb]
    \centering
        \begin{subfigure}{0.45\textwidth}
        \includegraphics[width=\textwidth]{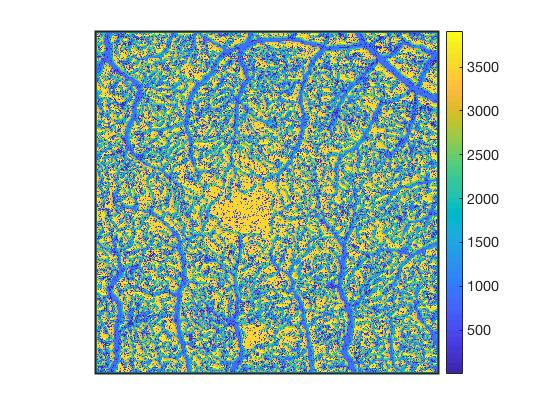}
        \caption{Assignment image}
        \label{fig:a}
    \end{subfigure}
    ~
    \begin{subfigure}{0.45\textwidth}
        \includegraphics[width=\textwidth]{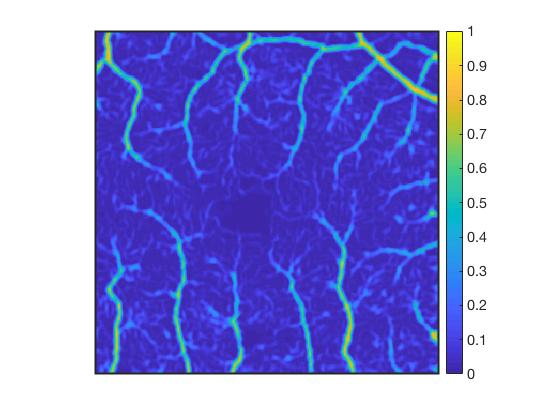}
        \caption{Probability image for larger vessels}
        \label{fig:p_v}
    \end{subfigure}
        \begin{subfigure}{0.45\textwidth}
        \includegraphics[width=\textwidth]{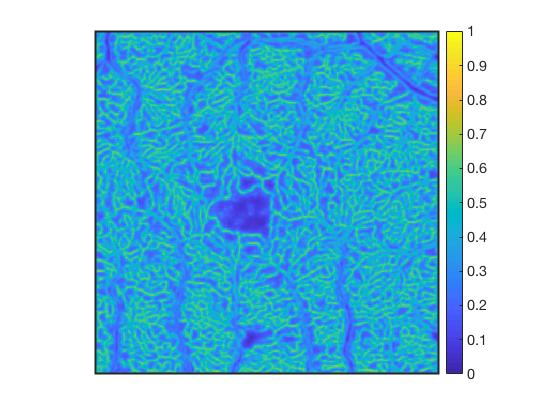}
        \caption{Probability image for capillaries}
        \label{fig:p_c}
    \end{subfigure}
        \begin{subfigure}{0.45\textwidth}
        \includegraphics[width=\textwidth]{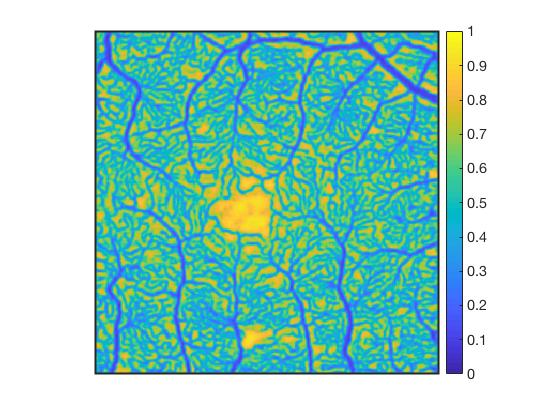}
        \caption{Probability image for background}
        \label{fig:p_b}
    \end{subfigure}
    \caption{Here, we show an assignment image in (a), where the color is the index of which dictionary element the patch around that pixel is assigned to. We also show the resulting probability images for the three different classes (b-d). 
    }
\end{figure}

The first step in processing the testing image is dictionary assignment. For this, we first extract a PCA feature vector from the patch around the pixel. Then we search the $k$-means tree to assign the pixel to the dictionary cluster. Our assignment is done by a simple greedy search through the $k$-means tree, and then assigning a feature vector to the nearest node in the tree. This does not guarantee that the feature vector is assigned to the closest node in the tree, but will generally ensure that similar patterns are grouped together. An illustration of the assignment is shown in \myfigref{fig:a}.

\subsubsection{Computing probability image.} 
From the dictionary assignment, we build a probability image $P$ of size $n \times m \times C$ by visiting all image pixels for each label, obtaining the probability information associated with their dictionary cluster, and adding it to $P$ in the spacial position corresponding to the pixel. Finally, we normalize $P$ such that the $C$ label probabilities sum to one. 

To capture the local appearance of the image, it can sometimes be advantageous to choose a slightly larger patch for computing the feature vector than the label patch, meaning that $N > M$. In that case, the $(N-M)/2$ boundary pixels will not be labeled. But in practice, this does not influence our analysis. However, $M$ and $N$ can be chosen independently such that it makes sense for the given segmentation problem.

The final step before obtaining the segmentation is to choose the most probable label from the $C$ labels.

\section{Experiments}
\begin{figure}[!tb]
    \centering
    \begin{subfigure}{0.45\textwidth}
        \includegraphics[width=\textwidth]{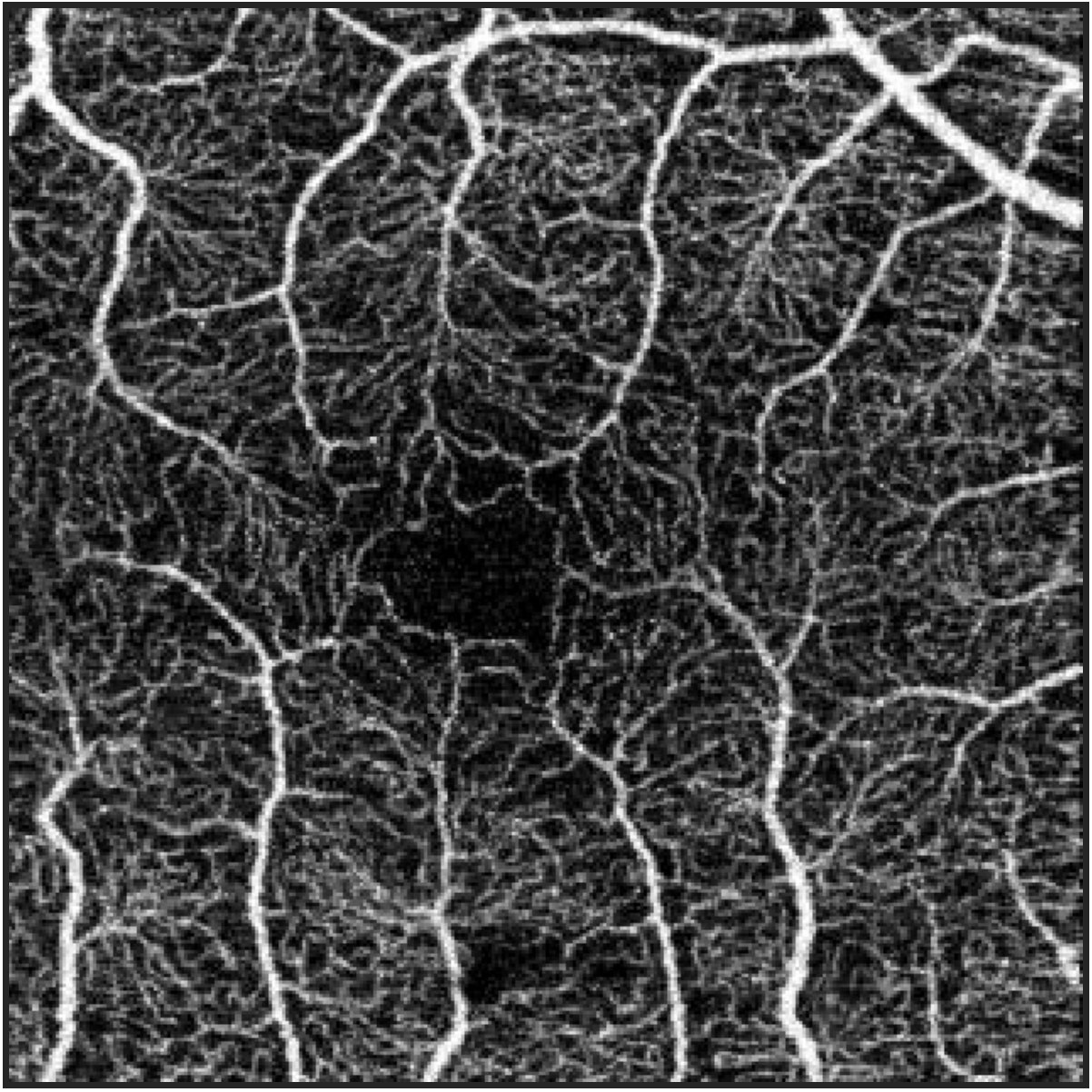}
        \caption{Training image}
        \label{fig:im_org}
    \end{subfigure}
    ~
    \begin{subfigure}{0.45\textwidth}
        \includegraphics[width=\textwidth]{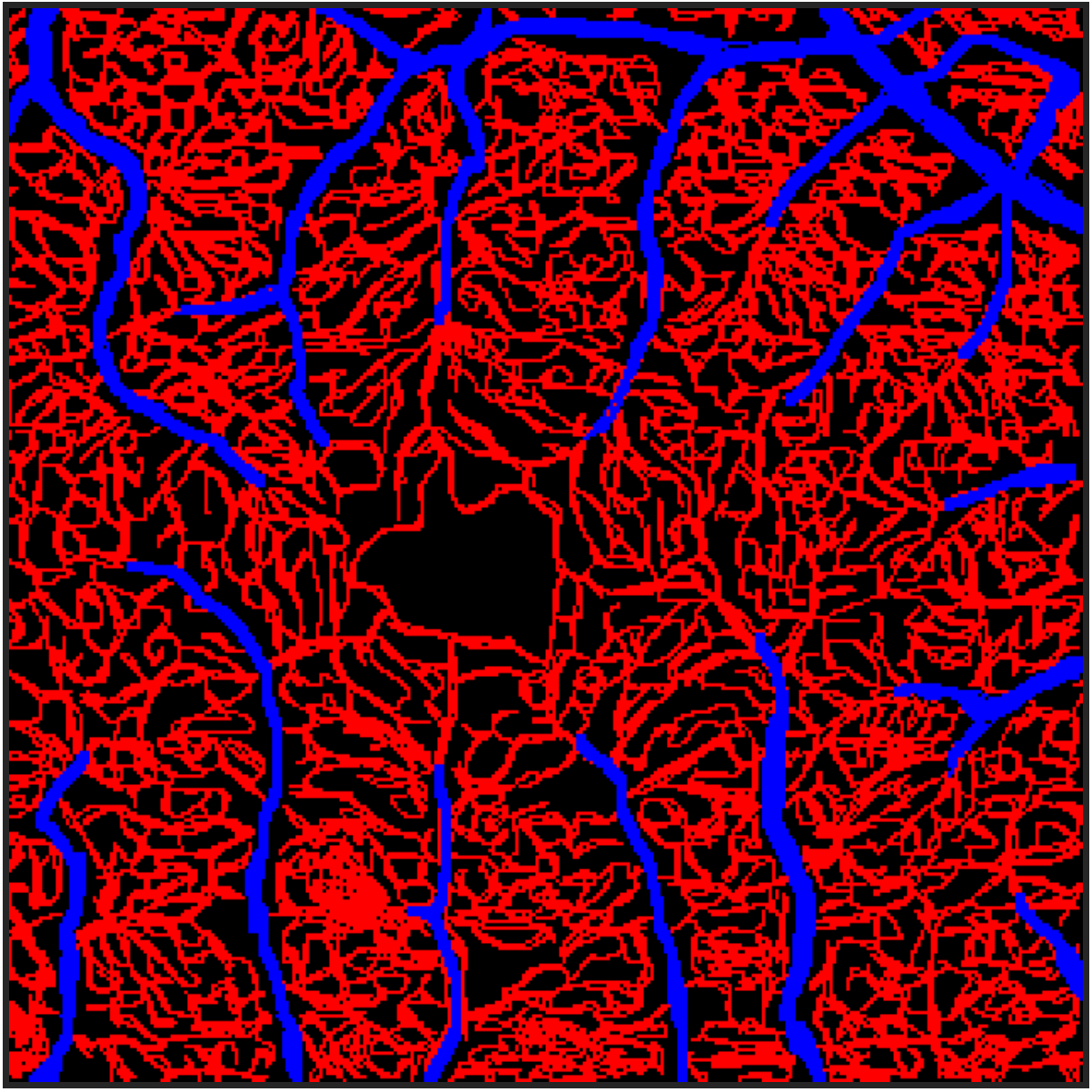}
        \caption{Manual labeling}
        \label{fig:im_org_gt}
    \end{subfigure}
    \caption{The dictionary is build from the training image seen in (a) and the corresponding manually labelled image shown in (b). Larger vessels are marked in blue, capillaries in red, and background in black.\label{fig:im_training}}
    \label{fig:gt}
\end{figure}

To train the model, we are using one OCTA image (a training image) with corresponding labeling, both shown in \myfigref{fig:im_training}. The OCTA image has been acquired by a swept source DRI OCT Triton, Topcon Medical Systems, Inc. The image was initially $320 \times 320$ pixels, but was upscaled by a factor of two to a final size of $640 \times 640$ pixels. Before using the image to train the dictionary, we applied an adaptive histogram equalization, which improved the contrast of the image \cite{pizer1987adaptive}. For adaptive histogram equalization the image is divided into patches of size $40 \times 40$ pixels with a contrast enhancement limit of 0.004 preventing oversaturation in homogeneous regions.

The three classes for describing the retinal microvasculature include capillaries, larger vessels (arterioles and venules), and background. Larger vessels are defined as vessels with a radius of at least twice the radius of the capillaries. Image patches for computing the PCA feature use $7 \times 7$ pixels ($N=7$). To enhance the variability of the patches, the features are extracted from both the original image, as well as a 90 degree rotation of the image. 50000 random patches ($K = 50000$)
are used to compute the PCA features where the $q = 10$ biggest components are used. 

Next, label patches of size $13 \times 13$ pixels are used for the probability dictionary ($M = 13$). The patch size is on the same scale as the capillaries that we wish to identify, and it has been determined through a parameter study. Figure \ref{fig:comp} shows the largest component of each of the six feature groups. Now, each non-boundary pixel in the image has six $10 \times 1$ vectors, which are concatenated into a $60$-dimensional feature vector. 

We then perform $k$-means clustering of $100000$ feature vectors by building a search tree with five layers and a branching factor of five (with a maximum of $3905$ clusters). We end up with $3905$ dictionary elements. The assignment image $A$ is created, see \myfigref{fig:a}, where the color corresponds to the index of the dictionary element for the image patch around that pixel and the corresponding probability images are shown in \myfigref{fig:p_v}-d.

To optimize the segmentation, a $3 \times 3$ weight matrix $\mathbf{W}$ is computed such that the resulting class probabilities of the training image equal the annotated class labels by
\begin{equation}
    \mathbf{W} = \min_\mathbf{W} \Vert \hat{L} - \hat{P}\mathbf{W}\Vert^2_2 \ , 
\end{equation}
where $\hat{L}$ is the label image arranged into an $nm \times C$ matrix, where each row contains the pixel-wise label probabilities, and $\hat{P}$ is the probability image arranged in the same way. This is solved as linear least squares problem.

\begin{figure}[!tb]
    \centering
    \begin{subfigure}{0.2\textwidth}
        \includegraphics[width=2.8cm,height=2.8cm]{Figures/12_segmentation_paper.png}
        \caption{SRL} 
    \end{subfigure}
    ~\hspace{0.3cm}
    \begin{subfigure}{0.2\textwidth}
        \includegraphics[width=3cm,height=2.8cm]{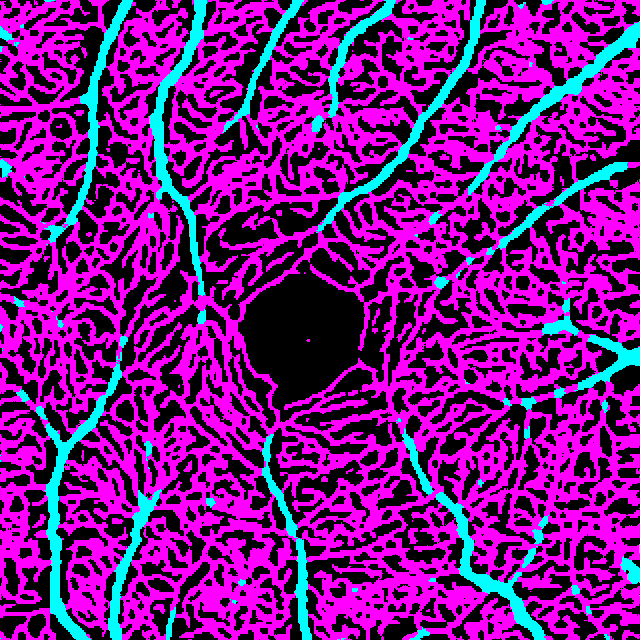}
        \caption{SRL output}
    \end{subfigure}
    ~\hspace{0.3cm}
    \begin{subfigure}{0.2\textwidth}
        \includegraphics[width=2.8cm,height=2.8cm]{Figures/15_segmentation_paper.png}
        \caption{DRL}
    \end{subfigure}
        ~\hspace{0.3cm}
    \begin{subfigure}{0.2\textwidth}
        \includegraphics[width=2.8cm,height=2.8cm]{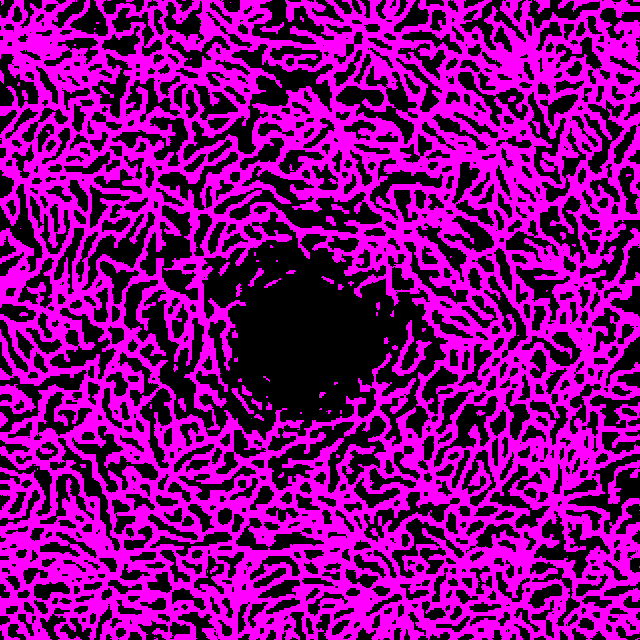}
        \caption{DRL output}
    \end{subfigure}
    ~
    \begin{subfigure}{0.2\textwidth}
        \includegraphics[width=2.8cm,height=2.8cm]{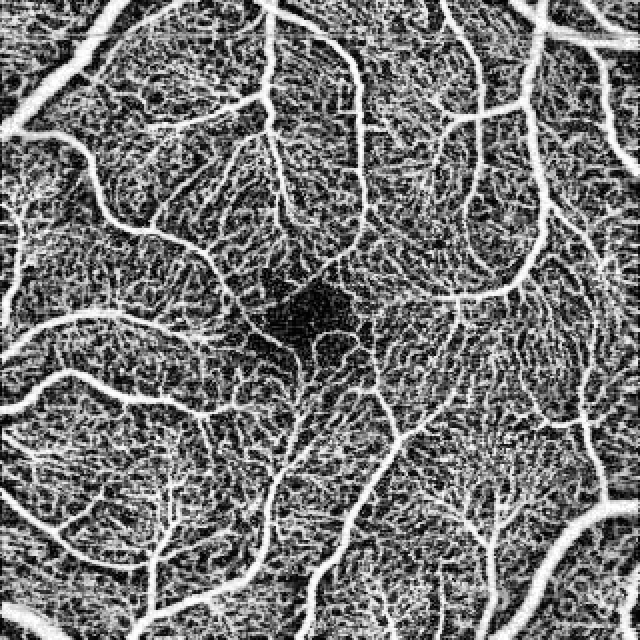}
        \caption{SRL}
    \end{subfigure}
        ~\hspace{0.3cm}
    \begin{subfigure}{0.2\textwidth}
        \includegraphics[width=3cm,height=2.8cm]{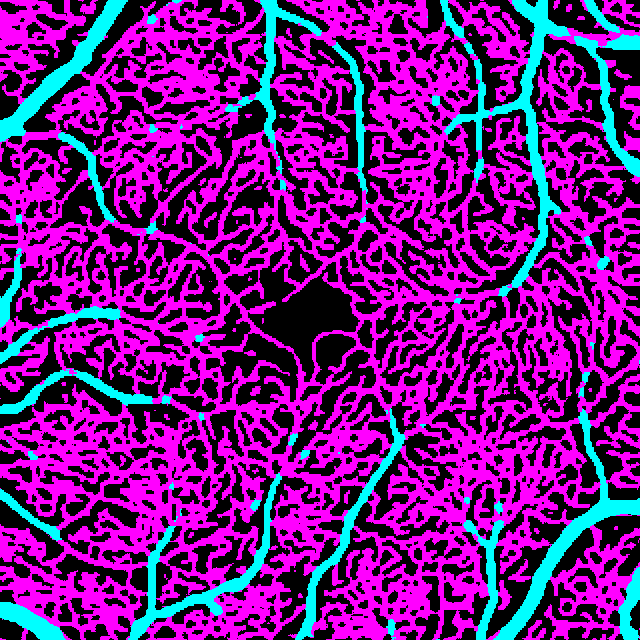}
        \caption{SRL output}
    \end{subfigure}
    ~\hspace{0.3cm}
    \begin{subfigure}{0.2\textwidth}
        \includegraphics[width=2.8cm,height=2.8cm]{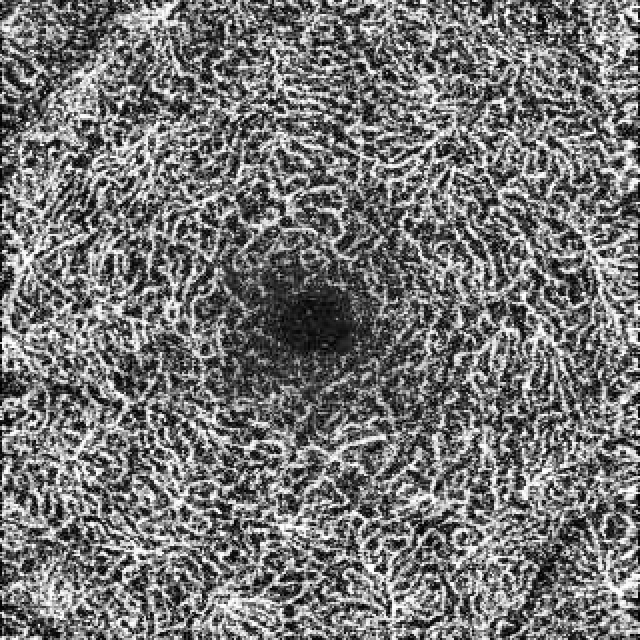}
        \caption{DRL}
    \end{subfigure}
        ~\hspace{0.3cm}
    \begin{subfigure}{0.2\textwidth}
        \includegraphics[width=2.8cm,height=2.8cm]{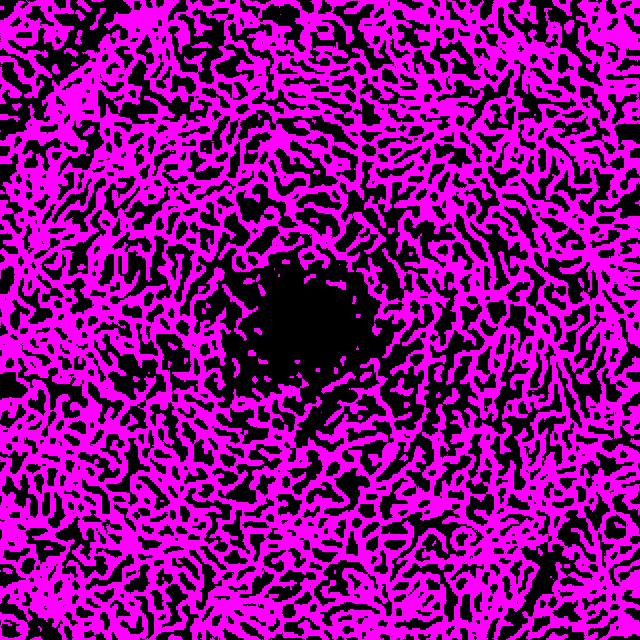}
        \caption{DRL output}
    \end{subfigure}
    ~
    \begin{subfigure}{0.2\textwidth}
        \includegraphics[width=2.8cm,height=2.8cm]{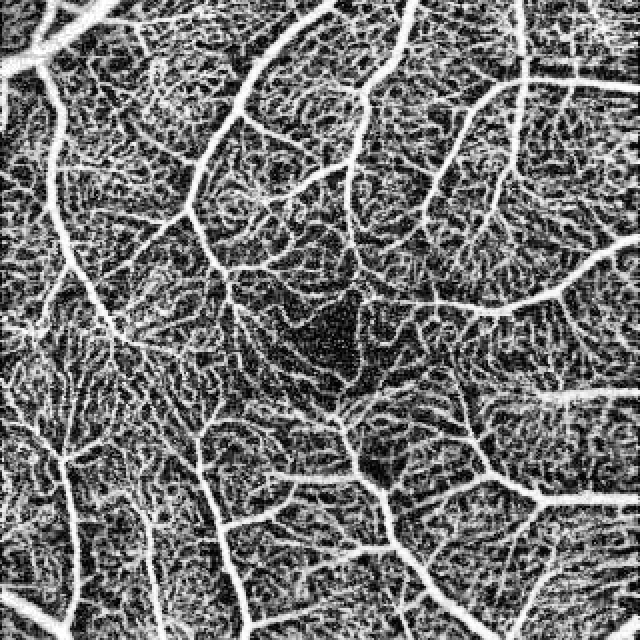}
        \caption{SRL} 
    \end{subfigure}
    ~\hspace{0.3cm}
    \begin{subfigure}{0.2\textwidth}
        \includegraphics[width=3cm,height=2.8cm]{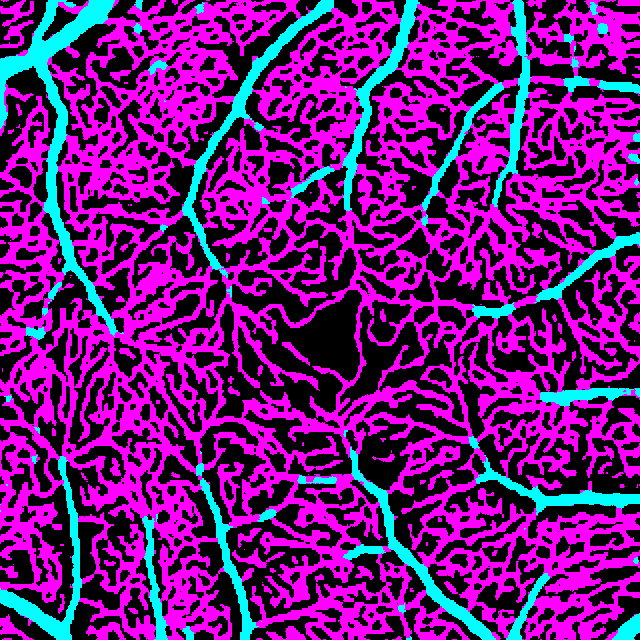}
        \caption{SRL output}
    \end{subfigure}
    ~\hspace{0.3cm}
    \begin{subfigure}{0.2\textwidth}
        \includegraphics[width=2.8cm,height=2.8cm]{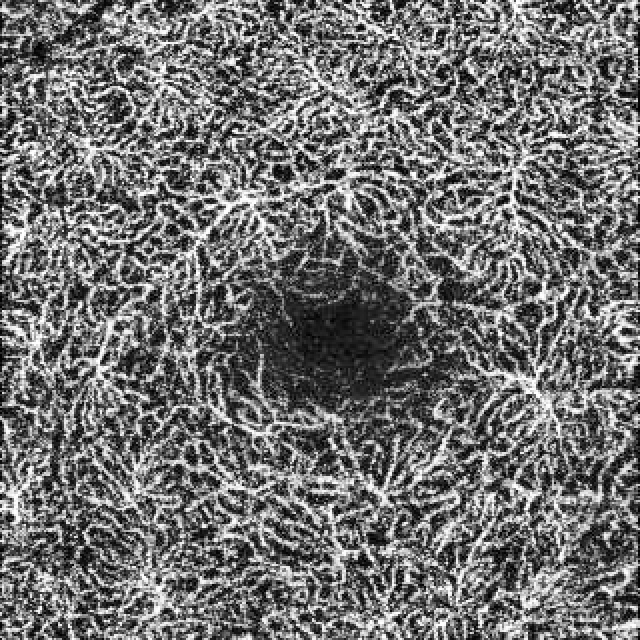}
        \caption{DRL}
    \end{subfigure}
        ~\hspace{0.3cm}
    \begin{subfigure}{0.2\textwidth}
        \includegraphics[width=2.8cm,height=2.8cm]{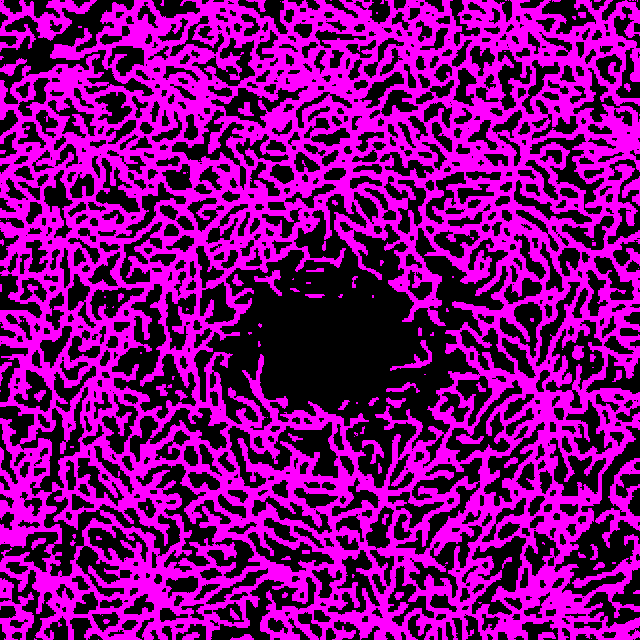}
        \caption{DRL output}
    \end{subfigure}
    ~
        \begin{subfigure}{0.2\textwidth}
        \includegraphics[width=2.8cm,height=2.8cm]{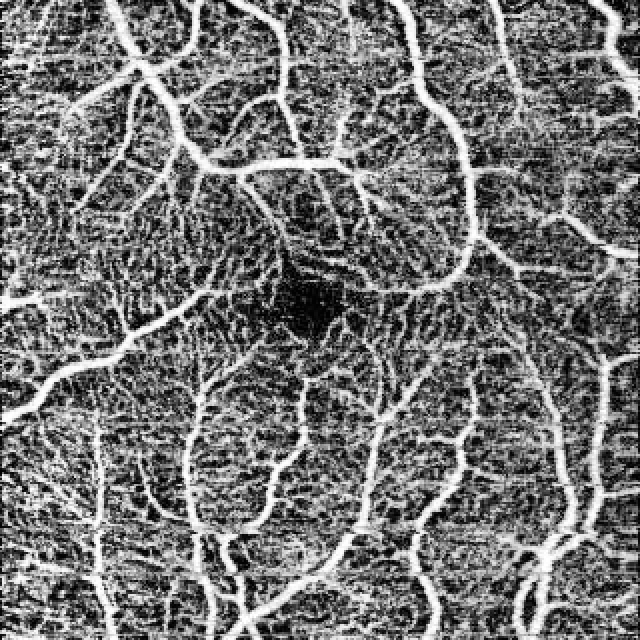}
        \caption{SRL} 
    \end{subfigure}
    ~\hspace{0.3cm}
    \begin{subfigure}{0.2\textwidth}
        \includegraphics[width=3cm,height=2.8cm]{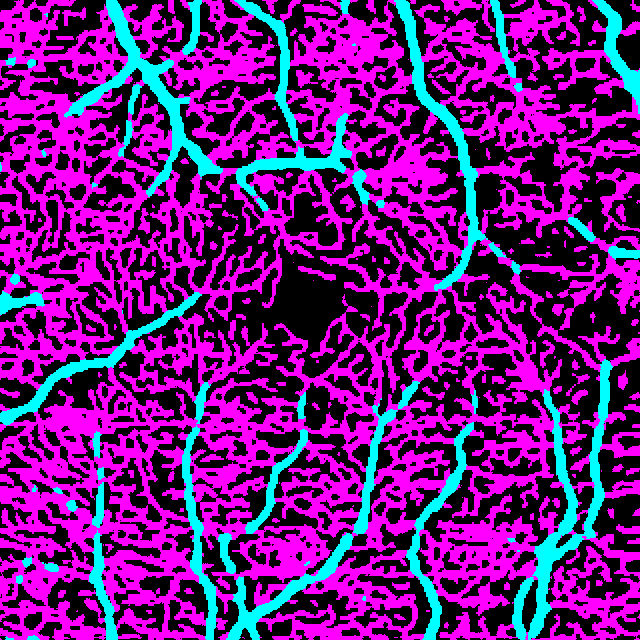}
        \caption{SRL output}
    \end{subfigure}
    ~\hspace{0.3cm}
    \begin{subfigure}{0.2\textwidth}
        \includegraphics[width=2.8cm,height=2.8cm]{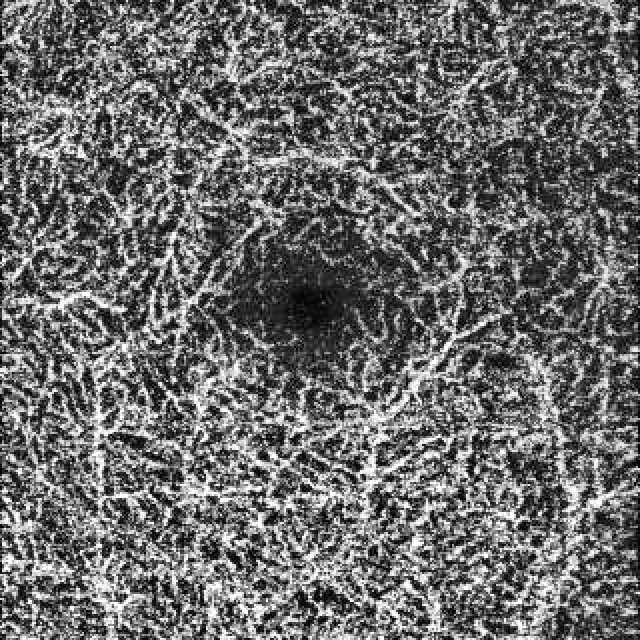}
        \caption{DRL}
    \end{subfigure}
        ~\hspace{0.3cm}
    \begin{subfigure}{0.2\textwidth}
        \includegraphics[width=2.8cm,height=2.8cm]{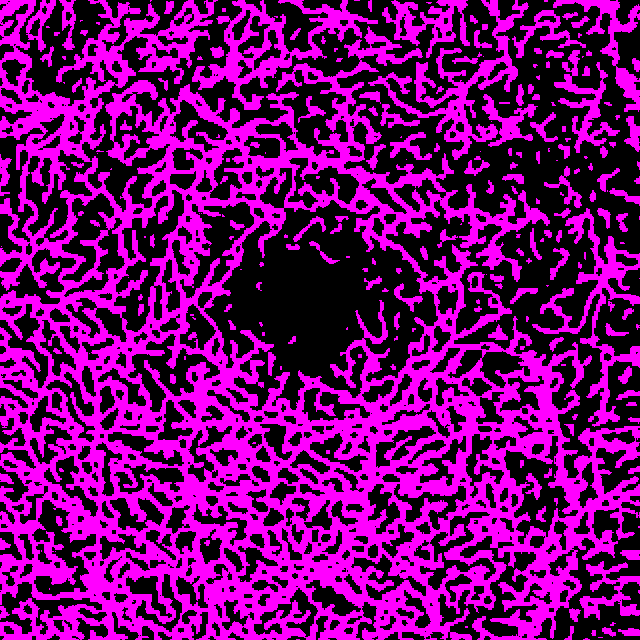}
        \caption{DRL output}
    \end{subfigure}
    ~
    \begin{subfigure}{0.2\textwidth}
        \includegraphics[width=2.8cm,height=2.8cm]{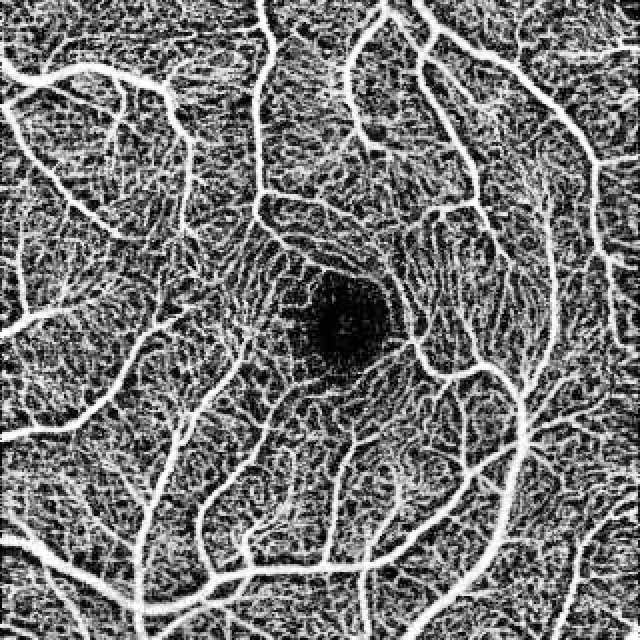}
        \caption{SRL} 
    \end{subfigure}
    ~\hspace{0.3cm}
    \begin{subfigure}{0.2\textwidth}
        \includegraphics[width=3cm,height=2.8cm]{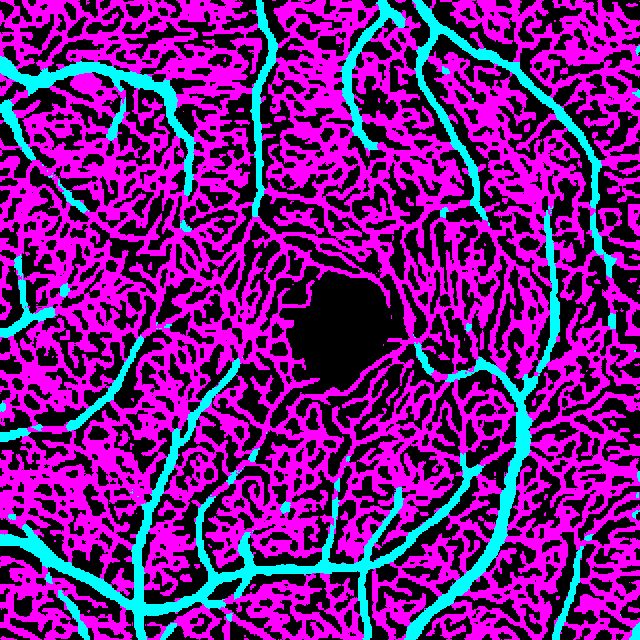}
        \caption{SRL output}
    \end{subfigure}
    ~\hspace{0.3cm}
    \begin{subfigure}{0.2\textwidth}
        \includegraphics[width=2.8cm,height=2.8cm]{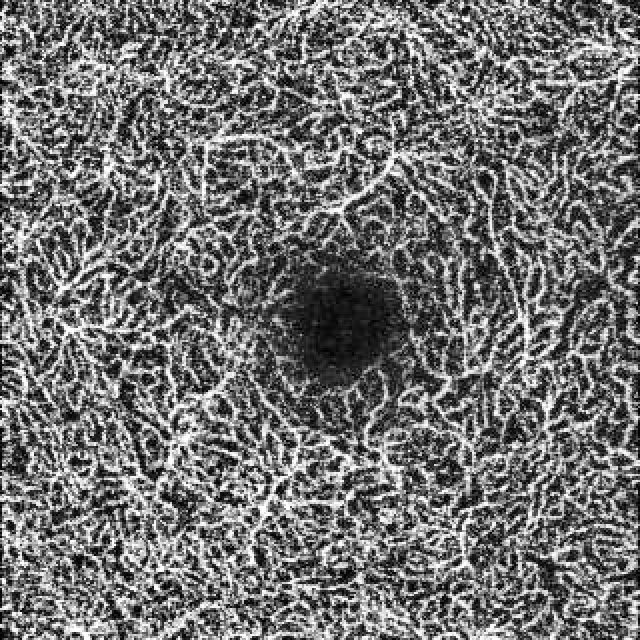}
        \caption{DRL}
    \end{subfigure}
        ~\hspace{0.3cm}
    \begin{subfigure}{0.2\textwidth}
        \includegraphics[width=2.8cm,height=2.8cm]{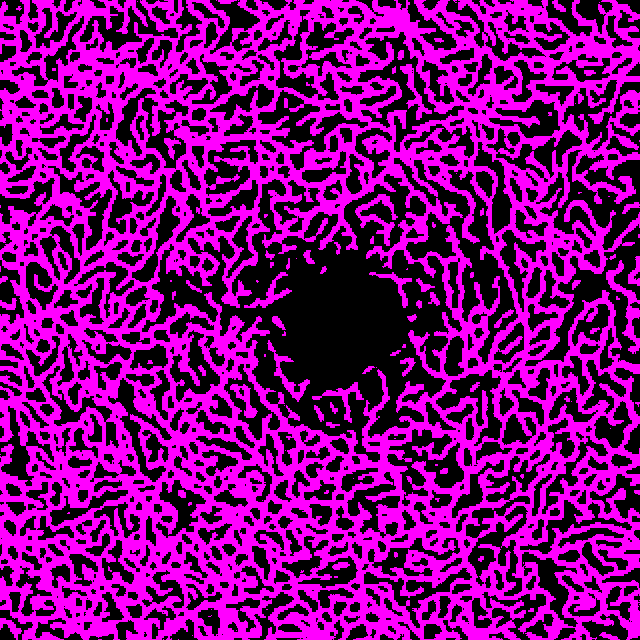}
        \caption{DRL output}
    \end{subfigure}    
    \caption{Examples of segmentation output. Subfigures (a)-(d) show the segmentation of subject 1, (e)-(h) show subject 2, (i)-(l) show subject 3, (m)-(p) show subject 4, and (q)-(t) show subject 5.}
    \label{fig:examples}
\end{figure}

The OCTA images, that we have worked with, consist of both the superficial retinal layer (SRL) and the deep retinal layer (DRL). Only the SRL is used to create the dictionary, as we wish to have training information containing larger vessels. Since there are mainly capillaries present in the DRL, the detected capillaries and the detected larger vessels are combined into one class in this layer. \myfigref{fig:examples} shows some examples of the resulting segmentation in both the superficial and deep retinal layers. It should be noted that these segmentations are obtained from a model that has been trained using one single annotated training image. It is very time consuming to annotate the detailed microvascular structures, and therefore it is advantageous that only one image is needed to obtain this result.

The segmentation model is relatively fast to train and run. Computing the PCA feature model takes around $4.25$ seconds, building the dictionary takes around $7.01$ seconds, and computing the segmentation from an unseen image using the trained models takes around $5.95$ seconds. The most time consuming part is the manual annotation of the training image. 

\section{Conclusion}
We have presented our dictionary-based segmentation method that allows segmentation of the retinal microvasculature from OCTA images. From a single annotated image, we obtain a clear separation into the three classes of capillaries, larger vessels, and background. The model is fast to compute, and gives an accurate separation of the three classes, allowing for quantitative assessment of the retinal microvasculature.

\clearpage
%
%
\bibliographystyle{splncs04}
\bibliography{mybibliography}

\begin{thebibliography}{10}
\providecommand{\url}[1]{\texttt{#1}}
\providecommand{\urlprefix}{URL }
\providecommand{\doi}[1]{https://doi.org/#1}

\bibitem{chu2016quantitative}
Chu, Z., et~al.: Quantitative assessment of the retinal microvasculature using
  optical coherence tomography angiography. J. Biomed. Opt  \textbf{21}(6),
  066008 (2016)

\bibitem{dahl2011learning}
Dahl, A., Larsen, R.: Learning dictionaries of discriminative image patches.
  In: Proc. BMVC. pp. 77.1--77.11. BMVA Press (2011)

\bibitem{dahl2014dictionary}
Dahl, A.B., Dahl, V.A.: Dictionary snakes. In: Pattern Recognition (ICPR), 2014
  22nd International Conference on. pp. 142--147. IEEE (2014)

\bibitem{dahl2015dictionary}
Dahl, A.B., Dahl, V.A.: Dictionary based image segmentation. In: Scandinavian
  Conference on Image Analysis. pp. 26--37. Springer (2015)

\bibitem{dahl2018content}
Dahl, V.A., Trinderup, C.H., Emerson, M.J., Dahl, A.B.: Content-based
  propagation of user markings for interactive segmentation of patterned
  images. arXiv preprint arXiv:1809.02226  (2018)

\bibitem{deng_measurements_2018}
Deng, W.: Measurements of retinal microvasculature in mice and humans with deep
  learning. {PhD}, University of Iowa, Iowa City, Iowa, USA (Aug 2018).
  \doi{10.17077/etd.84ques5w}

\bibitem{elad2010sparse}
Elad, M.: Sparse and redundant representations: from theory to applications in
  signal and image processing. Springer Science \& Business Media (2010)

\bibitem{eladawi_automatic_2017}
Eladawi, N., Elmogy, M., Helmy, O., Aboelfetouh, A., Riad, A., Sandhu, H.,
  Schaal, S., El-Baz, A.: Automatic blood vessels segmentation based on
  different retinal maps from {OCTA} scans. Computers in Biology and Medicine
  \textbf{89},  150--161 (Oct 2017). \doi{10.1016/j.compbiomed.2017.08.008}

\bibitem{engberg_automated_2020}
Engberg, A., Amini, A., Willerslev, A., Larsen, M., Sander, B., Kessel, L.,
  Dahl, A.B., Dahl, V.A.: Automated {Quantification} of {Macular} {Vasculature}
  {Changes} from {OCTA} {Images} of {Hematologic} {Patients}. In: Prooceedings
  of {International} {Symposium} of {Biomedical} {Imaging} (2020)

\bibitem{engberg2020mia}
Engberg, A.M.E., Erichsen, J.H., Christensen, A.N., Conradsen, K., Sander, B.,
  Kessel, L., Dahl, A.B., Dahl, V.A.: Quantifying changes in the macular
  vasculature after cataract surgery. Submitted to: Medical Image Analysis
  (2020)

\bibitem{engberg2019automated}
Engberg, A.M.E., Erichsen, J.H., Sander, B., Kessel, L., Dahl, A.B., Dahl,
  V.A.: Automated quantification of retinal microvasculature from {OCT}
  angiography using dictionary-based vessel segmentation. Communications in
  Computer and Information Science  (2019)

\bibitem{kashani2017optical}
Kashani, A.H., et~al.: Optical coherence tomography angiography: A
  comprehensive review of current methods and clinical applications. Prog Retin
  Eye Res.  \textbf{60},  66--100 (2017)

\bibitem{kim2016quantifying}
Kim, A.Y., et~al.: Quantifying microvascular density and morphology in diabetic
  retinopathy using spectral-domain optical coherence tomography angiography.
  Invest Ophthalmol Vis Sci  \textbf{57}(9),  OCT362--OCT370 (2016)

\bibitem{pizer1987adaptive}
Pizer, S.M., Amburn, E.P., Austin, J.D., Cromartie, R., Geselowitz, A., Greer,
  T., ter Haar~Romeny, B., Zimmerman, J.B., Zuiderveld, K.: Adaptive histogram
  equalization and its variations. Computer vision, graphics, and image
  processing  \textbf{39}(3),  355--368 (1987)

\bibitem{prentasic_segmentation_2016}
Prentašic, P., Heisler, M., Mammo, Z., Lee, S., Merkur, A., Navajas, E., Beg,
  M.F., Šarunic, M., Loncaric, S.: Segmentation of the foveal microvasculature
  using deep learning networks. Journal of Biomedical Optics  \textbf{21}(7),
  075008 (Jul 2016). \doi{10.1117/1.JBO.21.7.075008}

\bibitem{spaide2018optical}
Spaide, R.F., Fujimoto, J.G., Waheed, N.K., Sadda, S.R., Staurenghi, G.:
  Optical coherence tomography angiography. Progress in retinal and eye
  research  \textbf{64},  1--55 (2018)

\end{thebibliography}

\end{document}